\documentclass[fleqn,usenatbib]{mnras}
\usepackage{newtxtext,newtxmath}

\usepackage{epsfig}
\usepackage{physics}
\usepackage{amsmath,color,bm}
\usepackage{booktabs}
\usepackage{subfig}

\makeatletter
\renewcommand{\p@subfigure}{}
\makeatother
\usepackage{nameref}
\usepackage{graphicx}
\usepackage{graphics}
\usepackage[space]{grffile}
\usepackage{url}
\usepackage[utf8]{inputenc}
\usepackage{fancyref}
\usepackage{xcolor}
\usepackage{gensymb}

\usepackage[normalem]{ulem}
\usepackage{eqnarray,amsmath}

\newcommand{\onepluszextra}{1 + z_\mathrm{exc}}

\title[Detecting BNS post-merger in SMBH environments]{Can a binary neutron star merger in the vicinity of a supermassive black hole enable a detection of a post-merger gravitational wave signal?}

\author[Vijaykumar, Kapadia, Ajith]{Aditya Vijaykumar,$^1$
  Shasvath J. Kapadia,$^1$
  Parameswaran Ajith$^{1,2}$\\
  $^1$International Centre for Theoretical Sciences, Tata Institute of Fundamental Research, Bangalore 560089, India\\
  $^2$Canadian Institute for Advanced Research, CIFAR Azrieli Global Scholar, MaRS Centre, West Tower, 661 University Ave, Toronto, ON M5G 1M1, Canada}

\begin{document}
\maketitle

\begin{abstract}
The postmerger gravitational-wave (GW) signal of a binary neutron star (BNS) merger is expected to contain valuable information that could shed light on the equation of state (EOS) of NSs, the properties of the matter produced during the merger, as well as the nature of any potential intermediate merger product such as hypermassive or supramassive NSs.  However, the postmerger lies in the high frequency regime ($ \gtrsim 1000 $ Hz) where current LIGO-Virgo detectors are insensitive.  While proposed detectors such as NEMO, Cosmic Explorer and Einstein Telescope could potentially detect the postmerger for BNSs within $\mathcal{O}(10~\mathrm{Mpc})$, such events are likely to be rare. In this work, we speculate on the possibility of detecting the postmerger from BNSs coalescing in the vicinity of supermassive black holes (SMBH). The redshift produced by the gravitational field of the SMBH, as well as the BNS's proper motion around the SMBH, could effectively ``stretch'' the postmerger signal into the band of the detectors. We demonstrate, using a phenomenological model, that such BNS coalescences would enable constraints on the peak of the postmerger signal that would otherwise have not been possible, provided the degree of redshifting due to the SMBH can be independently acquired. We further show how such mergers would improve EOS model selection using the postmerger signal.  We discuss the mechanisms that might deliver such events, and the limitations of this work.
\end{abstract}

\section{Introduction}\label{sec:introduction}
	Understanding and constraining the neutron star (NS) equation of state (EOS) with observational data is arguably one of the most exciting and high-profile targets of astrophysical research \citep[for a review see, e.g.][]{Ozel2016}. Not only does it promise to shed light on the properties and structure of the neutron star, but it also enables a probe of matter at its most extreme densities, currently irreplicable in human-made laboratories \citep[for a review see, e.g. ][]{Lattimer2012}.  Electromagnetic (EM) observations of NSs can provide invaluable data on the NS-EOS, either via detections of periodic radio pulses from spinning NSs \citep[see, e.g.][]{Lattimer1990}, or via highly sensitive spectroscopy of soft X-rays carried out by the Neutron Star Interior Composition Explorer (NICER) \citep{NICERa, NICERb}.

Nevertheless, perhaps the most direct probe of the NS EOS would be enabled by gravitational-wave (GW) observations of compact binary coalescences (CBCs) containing at least one NS. The GWs from their inspirals could contain imprints of the degree of tidal deformation of the NS(s), which in turn would allow constraints on the NS EOS  \citep[see, e.g.][and references therein]{Chatziioannou2020}. Furthermore, the shape of the BNS postmerger signal could also enable an EOS model selection, since, for example, the location of the postmerger peak frequency is EOS dependent \citep{Clark2016}. The postmerger signal could also contain information on the evolutionary path of the merged binary; depending on the mass of the compact binary and the EOS of the NS, the merged object could either be a black hole (BH) or a heavy (hypermassive or supramassive) unstable NS \citep{HotokezakaHNS} which subsequently collapses to a BH after a period of time (which is also EOS dependent) \citep[see, e.g.][and references therein]{sarin2021}.

Detecting the postmerger signal therefore promises to open a treasure-trove of astrophysical riches. Unfortunately, the current sensitivity band of the LIGO-Virgo detectors does not encompass the high frequency regime necessary to detect the postmerger signal \citep{advligo, advvirgo}.  Thus, even though this GW detector network has detected over ninety CBCs (across three observing runs) of which a handful have been BNS detections \citep{GWTC-1, GWTC-2, LIGOScientific:2021djp, Olsen:2022pin, Nitz:2021zwj}, the postmerger signal remains hidden deep within noise. In fact, even GW170817, the very first and loudest BNS merger \citep{GW170817-DETECTION}, did not contain a postmerger signal that was detectable to the LIGO-Virgo network \citep{GW170817-PM}.  Furthermore, future planned upgrades to this network (A+ \citep{aplus_sensitivity}, Voyager \citep{Voyager_PSD}) will also be insensitive to the postmergers beyond $\mathcal{O}(1-10 \mathrm{Mpc})$ \citep{Clark2016}.

Arguably, the best bet to observe a postmerger signal in the next decade is to construct a high-frequency detector such as the proposed Neutron Star Extreme Matter Observatory---or NEMO \citep{NEMO} for short---even at the expense of a loss of sensitivity at the lower frequency end. The lower frequency piece of the GW would be captured by the other (LIGO-Virgo-Kagra) detectors in the network. However, the concept for NEMO-like detectors are in the early stage of their development.  Alternatively, a third-generation detector network consisting of two Cosmic Explorers \citep{CE} and one Einstein Telescope \citep{ET} could also detect the postmerger.  Nevertheless, these detectors may struggle to detect the postmerger signals coming from distances of $\mathcal{O}(10-100~\mathrm{Mpc})$ or farther \citep{Clark2016, sarin2021}. 

In this work, we consider a scenario in which a postmerger signal would be detectable, even in the absence of a high-frequency detector.  If a BNS merger were to occur in the vicinity of a supermassive black hole (SMBH), then the gravitational redshift produced by the potential of the SMBH, as well as the Doppler redshift due to the BNSs proper motion around the SMBH (assuming it has a velocity component pointing radially away from the Earth), could effectively ``stretch'' the postmerger signal into the band of the detectors.  At least two mechanisms have been proposed in the literature to deliver a compact binary merger in the vicinity of an SMBH, even near its innermost stable circular orbit (ISCO). One involves the existence of migration traps \citep{Peng2021}, while the other pertains to tidal capture \citep{ChenLiCao2019, ChenHan2018}. 

Nuclear star clusters are expected to be sites where compact binary mergers occur in significant numbers. The mechanisms that drive these mergers in the vicinity of the SMBH include mass segregation \citep{Miralda2000, Freitag2006},  tidal capture \citep{ChenHan2018}, and tidal perturbation \citep{Antonini2012}. Among these mechanisms, tidal capture could enable mergers close to the SMBH, even within $10 R_{s}$, where $R_{s}$ is the gravitational (Schwarzschild) radius of the SMBH. Some authors have even argued that some of the heavier masses observed in LIGO-Virgo's BBHs can be attributed to gravitational and Doppler redshifting suffered by their GWs, since they could have merged in the vicinity of SMBHs via tidal capture \citep{ChenLiCao2019}.

Several studies have also suggested that the accretion disk in an active galactic nucleus (AGN) could contain a large number of compact objects. Some of them could have been snared by the accretion disk due to repeated collisions with it \citep{Norman1983, Syer1991, Artymowicz1993, MacLeod2020}, while others could have become compact objects from massive stars residing in the outer parts of the disk \citep{Goodman2004, Levin2007, Wang2010}.  

These compact objects could then form binaries, and merge within Hubble time, due to interactions with each other or with the gas-rich environment. On the other hand, the interaction of the binary with the surrounding gas would cause it to migrate towards the central SMBH powering the AGN. However, opposing wind-driven torques near the ISCO of the SMBH produces a migration trap, where the binary could live out the remainder of its life and merge. \citep{Peng2021}

AGNs have been seriously considered in the literature as potential sites for the mergers detected by LIGO-Virgo. Recent work suggests that, under certain assumptions, as many as 50\% of the BBHs detected in the third observing run (O3) could be associated with AGN disks \citep{McKernan2020}. The SMBHs driving the AGNs enable mass segregation \citep{Freitag2006} that allow stellar mass BBHs to merge in the vicinity of the SMBHs, while their heavier masses make them less likely to acquire the necessary escape velocity to be ejected (kicked) out of the AGN.  

On the other hand, BNS mergers in AGN disks are expected to contribute a smaller fraction to the total BNS merger rate than the corresponding contribution of BBH mergers to the total BBH merger rate. AGNs are therefore not expected to be the dominant sites for LIGO-Virgo's BNS mergers. This can be attributed at least in part to mass-segregation pushing NSs away from the SMBHs, and their lighter masses allowing them to acquire the necessary velocities to get ejected out of the AGN. Nevertheless, the BNS merger rate within AGNs could be up to $\sim 4$ times as large as the BBH merger rate \citep{McKernan2020}.  From \citet{Ford2021}, the BBH merger rate in AGNs can be as high as $\sim 20~\mathrm{Gpc}^{-3}\mathrm{yr}^{-1}$, which places a BNS merger rate upper limit of $\sim 80~\mathrm{Gpc}^{-3}\mathrm{yr}^{-1}$. Optimistically, then, in a 10-year 3G era, $\mathcal{O}(1000)$ BNS mergers in AGNs could be detected within $1000$ Mpc.  Interaction of the BNSs with the gas-rich environment,  and mechanisms such as tidal capture, could then cause a small fraction of them to merge in the vicinity of the SMBH, even inside the last migration trap close to the ISCO of the SMBH \citep{Peng2021}.

It is therefore reasonable to speculate that a handful ($\mathcal{O}(1-10)$) of BNSs could merge close to SMBHs, and their GWs redshifted sufficiently to present a detectable postmerger signal in a ten-year 3G era.  We show that such a GW event would allow a significantly improved postmerger signal-to-noise ratio (SNR), constraint on the postmerger peak frequency (provided the degree of redshifting could be acquired using independent means), and EOS model selection. 

The rest of the paper is organized as follows. Section~\ref{sec:method} delineates the redshifting of the detector frame mass and luminosity distance estimates, the computation of the postmerger SNR, a phenomenological model of the postmerger signal and inference of the peak frequency, and an approximate EOS (Bayesian) model selection scheme. Section~\ref{sec:results} summarizes the results, highlighting the improvements due to gravitational and Doppler redshifting. Section~\ref{sec:conclusion} discusses ways of inferring the true peak frequency from the redshifted peak frequency, the limitations of this work, and potential avenues for future work.

\section{Motivation and Methods}\label{sec:method}
\subsection{Redshifting of the GW signal}

\begin{figure*}
	\centering
	\includegraphics{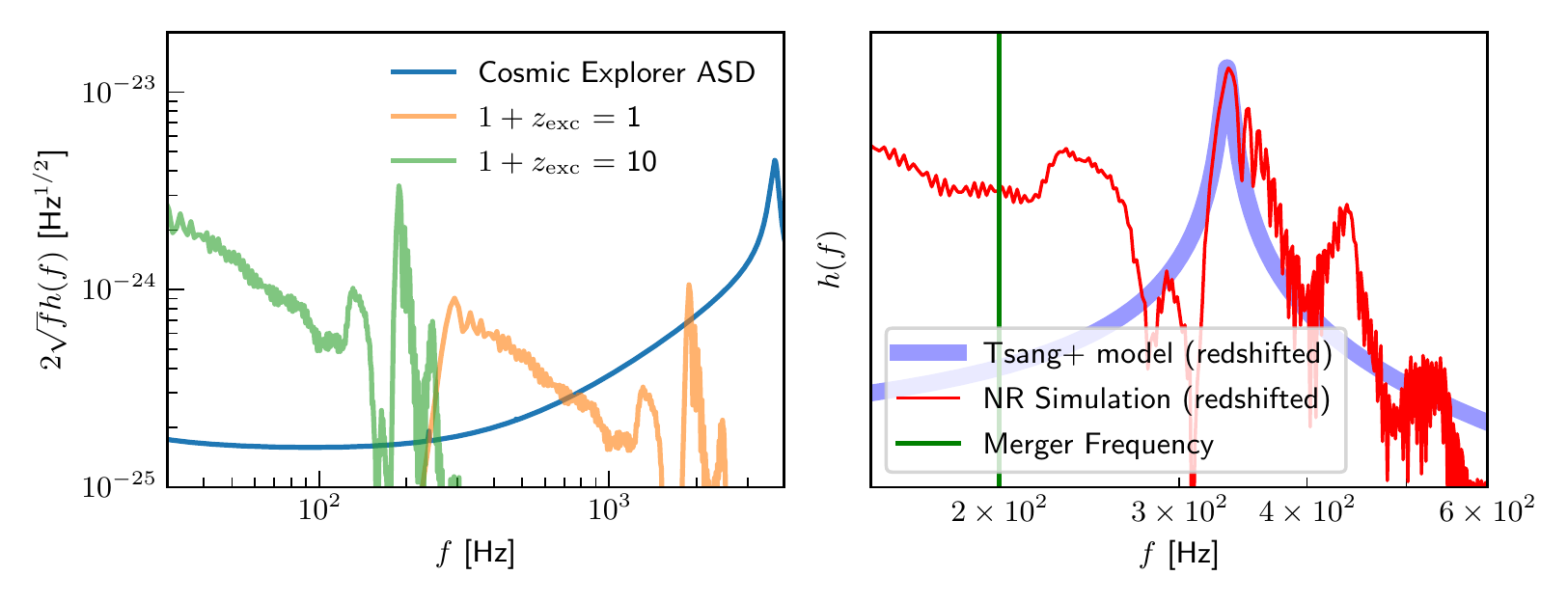}
	\caption{(Left) Illustration of the the redshifting and the ``dragging'' of the postmerger into the sensitive band of the detector. 
	We use the frequency-domain GW strain from the BAM:0098 numerical relativity simulation \citep{Bernuzzi:2014kca} of the Computational Relativity (CoRe) database \citep{CoRe}.
	The orange line corresponds to a BNS merger with $\onepluszextra=1 $ \textit{ie.} it incurs no extra redshift due to gravitational/doppler redshifting.
	We note that most of the postmerger signal is hidden under the expected noise amplitude spectral density (ASD) of the third-generation Cosmic Explorer detector (blue solid line). 
	However, when $\onepluszextra=10$ (green line), the signal gets pulled into band and most of the postmerger is now above the noise ASD. 
	(Right) Illustration of the match between the numerical simulation and the phenomenological model from \citet{Tsang2019}. The BAM:0098 waveform is plotted in the frequency domain (red line), assuming $ \onepluszextra = 10 $.
	We plot eq. \ref{eqn:pheno-model-redshifted} with $ f_\mathrm{peak} = 2000 $ Hz, $ f_\mathrm{spread} = 20 $ Hz and $ 1 + z = 10 $ (blue line).
	We also show the merger frequency calculated using an analytical expression in equation \ref{eqn:analytical-merger-frequency} (vertical green line) and use it to denote the start of the postmerger regime.
	The model captures the peak frequency and the spread in the frequency domain as expected, but fails to capture the other subdominant features.}
	\label{fig:redshift_illustration}
	
\end{figure*}

GWs from CBCs in the neighbourhood of SMBHs will undergo three distinct kinds of redshifts. These are: 

\begin{enumerate}
	\item the cosmological redshift  due to the expansion of the universe
	\item the Doppler redshift due to the motion of the BNS around the SMBH (if it has a velocity component that points radially away from the observer)
	\item the gravitational redshift due to the GWs climbing out of the SMBH's potential well,
\end{enumerate}

Each of these redshifts dilates the time at the location of the CBC, as measured by a distant observer.  As a result, the frequency of the GWs reaching the observer is reduced with respect to the frequency produced at the CBC's location. Since frequency is degenerate with the total mass of the system, the inferred component masses will be an overestimate of the true (source-frame) masses by an amount that depends on the degree of redshifting.

In general, computing the total redshift of such GWs is complicated, especially since it would require a single coordinate system to describe the metric in the vicinity of the SMBH as well as the metric of the expanding universe. Following \citet{ChenLiCao2019}, we approximate the total redshift to be a product of the cosmological redshift (estimated assuming standard cosmology \citep{Planck2018}), and an ``excess'' redshift due to the geodesic motion of a test particle in the Kerr metric \citep{Kerr1963} of the SMBH:
\begin{equation}
\label{eq:redshift_relation}
1 + z_{\mathrm{tot}} = (1 + z_{\mathrm{cos}})(1 + z_{\mathrm{exc}})
\end{equation}

Note that, we are concerned exclusively with the postmerger signal of a BNS whose wavelength (even after redshifting),  is several orders of magnitude smaller than the gravitational radius of the SMBH  ($M_{\mathrm{SMBH}} > 10^6 M_{\odot}$). We can therefore estimate the gravitational redshift in the short-wavelength limit (as with light) \citep{Isaacson1968}.\footnote{We assume $1 + z_{\mathrm{exc}}$ values to be $\leq 5$. Values greater than (or even close to) this value are likely to be quite rare, since such redshifts would require mergers close to the ISCO of highly spinning SMBHs.  We also neglect the effects of gravitational lensing, which might occur for mergers that happen behind the SMBH with respect to the observer.} The redshifting of the postmerger signal effectively ``drags'' the postmerger peak towards (or even into) the detector's sensitivity band, thus enabling its detection. This is illustrated in Figure~\ref{fig:redshift_illustration} (left panel) for an extreme value of the excess redshift.

The inferred redshifted mass (say the chirp-mass), will be changed from its source-frame value to its detector frame value, by $\mathcal{M} \rightarrow \mathcal{M}(1 + z_{\mathrm{tot}})$ \citep{Schutz1986, ChenLiCao2019}.  The inferred comoving distance is also biased by the redshifting. With cosmological redshift, this bias provides the luminosity distance $d_{L}$ \citep{HolzHughes2005}. The additional redshift due to gravity and the BNS's proper motion, then results in a biased estimate of the luminosity distance: $d_L \rightarrow d_L (1 + z_{\mathrm{exc}})$ \citep{ChenLiCao2019}.

\subsection{Postmerger SNR}
\label{subsec:postmerger-snr}
The prevalent method to search for GW signals of known shape in detector data is matched-filtering \citep{Wainstein1970},  where a GW template is cross-correlated with the data to produce a signal-to-noise ratio (SNR) statistic.  If the detector noise is both stationary and Gaussian,  and assuming that the template is a perfect representation of the signal in the data, then the optimal SNR is given by \citep[see, e.g.][]{creighton2012}:
\begin{equation}\label{eq:snr}
\rho = \sqrt{4\int_{0}^{\infty}\frac{\left | \tilde{h}(f) \right|^2}{S_n(f)}df}
\end{equation}
where $\tilde{h}(f)$ is the Fourier transform of the time-domain GW strain template $h(t)$, and $h(t)$ is a linear combination of the two GW polarizations, with coefficients that correspond to the response of the detector to each of the polarizations (often called the antenna pattern functions) \citep{schutz1987antenna, tinto1987antenna}. 

As mentioned earlier, this work restricts itself to the postmerger phase of the BNS merger. The postmerger phase is (in this work) defined to span frequencies $f \in \qty[f_{\mathrm{merg}} (1 + z_{\mathrm{tot}})^{-1},  f_{\mathrm{max}}]$.  We use the following analytical fit to calculate the merger frequency of binary neutron stars \citep{Breschi:2019srl}:
\begin{equation}
	\label{eqn:analytical-merger-frequency}
	M f_\mathrm{merg} = \qty(3.3184 \times 10^{-2})	 \dfrac{1 + (1.3067 \times 10^{-3}) \xi_{3199.8}}{1 + (5.0064 \times 10^{-3})\xi_{3199.8}} \qq{,}
\end{equation}
where $ \xi_{3199.8} = \kappa^T_2 + (3199.8)(1 - 4 \eta) $. Here, $\eta$ is the symmetric mass ratio $ \eta = \qty(m_1 m_2) / (m_1 + m_2)^2 $.  $M$ is the total source-frame mass of the binary, which, for this work, we assume to be (in units of $M_{\odot}$) $M = m_1 + m_2 = 1.35 + 1.35 = 2.7$. $ \kappa^T_2 $ is the effective tidal parameter for the binary at quadrupolar order in the EOB framework\footnote{For equal-mass binaries considered in this work, $\kappa^T_2$ is related to the oft-used $\tilde{\Lambda}$ parameter \citep{Flanagan:2007ix} by $\tilde{\Lambda} = \frac{16}{3} \kappa^T_2$ \citep{Bernuzzi:2014kca}.} \citep{Bernuzzi:2014kca}. We set $f_{\mathrm{max}} = 4000$ Hz. 

\subsection{A phenomenological postmerger model}
\label{subsec:pheno-model}

A finite number of numerical relativity waveforms that represent certain discrete points in the postmerger parameter space are currently available. However, a sufficiently accurate method to generate postmerger waveforms at arbitrary parameters does not exist at the time of writing.
For this work, we follow \citet{Tsang2019} to construct a simple phenomenological postmerger waveform that effectively mimics the postmerger peak frequency\footnote{There exist other models for the post-merger that capture more of its spectral features \citep{Easter:2018pqy, Soultanis:2021oia}. However, we restrict to this model for its functional simplicity.}.  Specifically, the waveform is modelled as a damped sinusoid in the time domain, which (Fourier) transforms to a Lorentzian.  We therefore adopt the following three-parameter ansatz for the frequency domain \citep{Tsang2019}:
\begin{equation}
\tilde{h}(f) = \frac{c_0 \ f_\mathrm{spread}}{\sqrt{(f - f_\mathrm{peak})^2 + f_\mathrm{spread}^2}}e^{-i\tan^{-1}\qty(\frac{f - f_\mathrm{peak}}{f_\mathrm{spread}})}
\label{eqn:pheno-model}
\end{equation}
The peak frequency is captured by $f_\mathrm{peak}$, the width of the postmerger is modelled by $f_\mathrm{spread}$, while $c_0$ is an overall amplitude scaling parameter.
Accounting for the redshift $ z_{\mathrm{tot}} $, this becomes:
\begin{equation}
\tilde{h}(f) = \frac{c_0 \ f_\mathrm{spread}}{\sqrt{\qty(f(1 + z_{\mathrm{tot}}) - f_\mathrm{peak})^2 + f_\mathrm{spread}^2}}e^{-i\tan^{-1}\qty(\frac{f(1 + z_{\mathrm{tot}}) - f_\mathrm{peak}}{f_\mathrm{spread}})}
\label{eqn:pheno-model-redshifted}
\end{equation}
The comparison of this model to a numerical simulation is illustrated in Figure \ref{fig:redshift_illustration} (right panel).
The model, for a certain choice of parameter values, captures the peak of the postmerger and the spread in frequency as expected.
However, it fails to capture other subdominant features in the postmerger.

\subsection{Bayesian parameter estimation }
The core of Bayesian parameter estimation is the estimation of the posterior distribution $ p(\vec{\theta} | d) $ of the parameters $ \vec{\theta} $, given the data $ s $ and a signal model $ h(\vec{\theta}) $.  Using Bayes theorem:
\begin{equation}
p(\vec{\theta}| s) = \frac{\pi(\vec{\theta}) p(s| \vec{\theta})}{p(s)} \qq{,}
\end{equation}
where $ \pi(\vec{\theta}) $ is the prior distribution, $  p(s| \vec{\theta}) $ is the likelihood, and $ p(s) := \int \pi (\vec{\theta}) p(s|\vec{\theta}) \dd{\vec{\theta}} $ is the marginalized likelihood, or the \textit{evidence}. If the noise is stationary Gaussian, the likelihood can be written as \citep{cutler1994}:
\begin{equation}
\mathcal{L} \equiv p(s | \vec{\theta}) \propto e^{-\frac{1}{2}(s-h(\theta) \mid s-h(\theta))} \qq{,}
\end{equation}
where $s$ is the detector strain data, and $(\cdot \mid \cdot)$ represents the noise-weighted inner product:
\begin{equation}\label{eq:innerproduct}
(a \mid b) = 2\int_{f_{\mathrm{min}}}^{f_{\mathrm{max}}}\frac{\tilde{a}(f)\tilde{b}^*(f)}{S_n(f)}df
\end{equation}
Here, $\qty[f_{\mathrm{min}}, f_{\mathrm{max}}]$ is the frequency range over which the inner product is to be evaluated. 

It would be germane to mention here that the prior, likelihood, evidence and the posterior, are implicitly conditioned on the assumed physical model used to produce the postmerger, in particular, the NS EOS. For notational simplicity, this has been omitted, but will be explicitly mentioned in the next subsection.

\subsection{Equation of State Model Selection}
\label{subsec:eos-model-selection}
A number of NS EOSs have been proposed in the literature. While some of the harder EOSs have been disfavoured by analyses of GW170817 \citep{GW170817-EOS}, what is the true NS EOS still remains an open question. BNS postmerger detections have the potential to further rule out candidate EOSs. 

A comparative analysis that quantitatively determines if one EOS model (say,  EOS1) is favored over another EOS model (say, EOS2), involves evaluating a Bayes factor. This factor is simply the ratio of the marginal likelihoods (evidences) under each of two hypotheses - the hypothesis $\mathcal{H}_1 (\mathcal{H}_2)$ that the signal in the data known to be a BNS postmerger corresponds to EOS1 (EOS2).  The Bayes factor can be evaluated as follows:
\begin{equation}
\mathcal{B}^1_2 = \frac{p(s\mid \mathcal{H}_1)}{p(s\mid \mathcal{H}_2)} = \frac{\int p(\vec{\theta}\mid\mathcal{H}_1)p(s\mid \vec{\theta}, \mathcal{H}_1)d\vec{\theta}}{\int p(\vec{\theta}\mid\mathcal{H}_2)p(s\mid \vec{\theta}, \mathcal{H}_2)d\vec{\theta}}
\end{equation}
In the absence of postmerger waveform models (for different EOSs) that can smoothly span the domain of the BNS's intrinsic parameters, we resort to a well-known approximation to this Bayes factor.  In particular, given that the data $s$ contains a postmerger signal corresponding to EOS1 with intrinsic parameters $\vec{\theta} = \vec{\theta}_1$,  this approximation can be evaluated using the optimal SNR and the fitting factor (FF)\footnote{The extrinsic parameters are semi-independent of the intrinsic parameters, and are assumed to be known for simplicity.}\citep{Cornish2011, Vallisneri2012}
\begin{equation}
\ln\mathcal{B}^1_2 \approx (1 - \mathrm{FF})\rho^2
\end{equation}
where the FF is defined as the noise-weighted inner product (cf. Eq.~\ref{eq:innerproduct}) of the waveforms under each of the two hypotheses, maximized over the intrinsic parameters $\vec{\theta}$:
\begin{equation}
\mathrm{FF} \equiv \left(h_{\mathrm{EOS1}}(\vec{\theta}_1)\mid h_{\mathrm{EOS2}} (\vec{\theta})\right)_{\mathrm{max}(\vec{\theta})}
\end{equation}
and $\rho$ is evaluated using $h_{\mathrm{EOS1}}(\vec{\theta}_1)$ (cf. Eq.~\ref{eq:snr}).

\section{Results}\label{sec:results}
	\begin{figure*}

	\includegraphics[width=\textwidth]{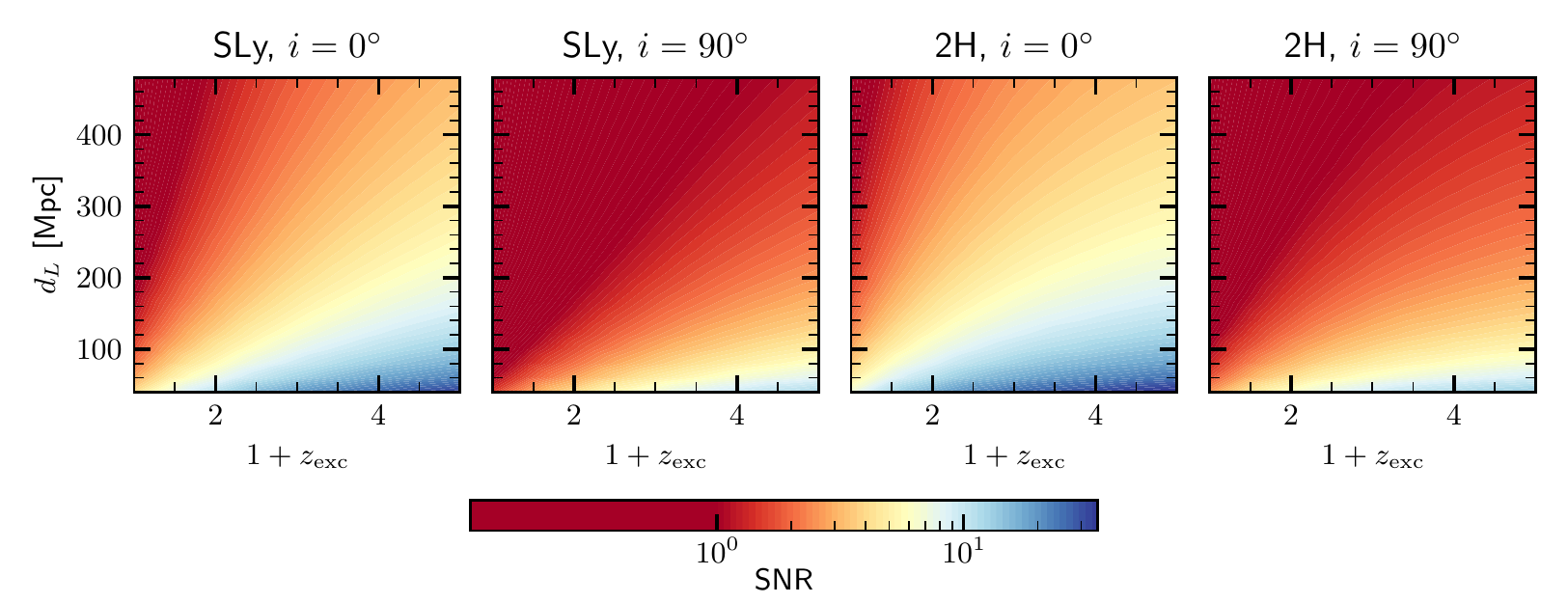}
	\includegraphics[width=\textwidth]{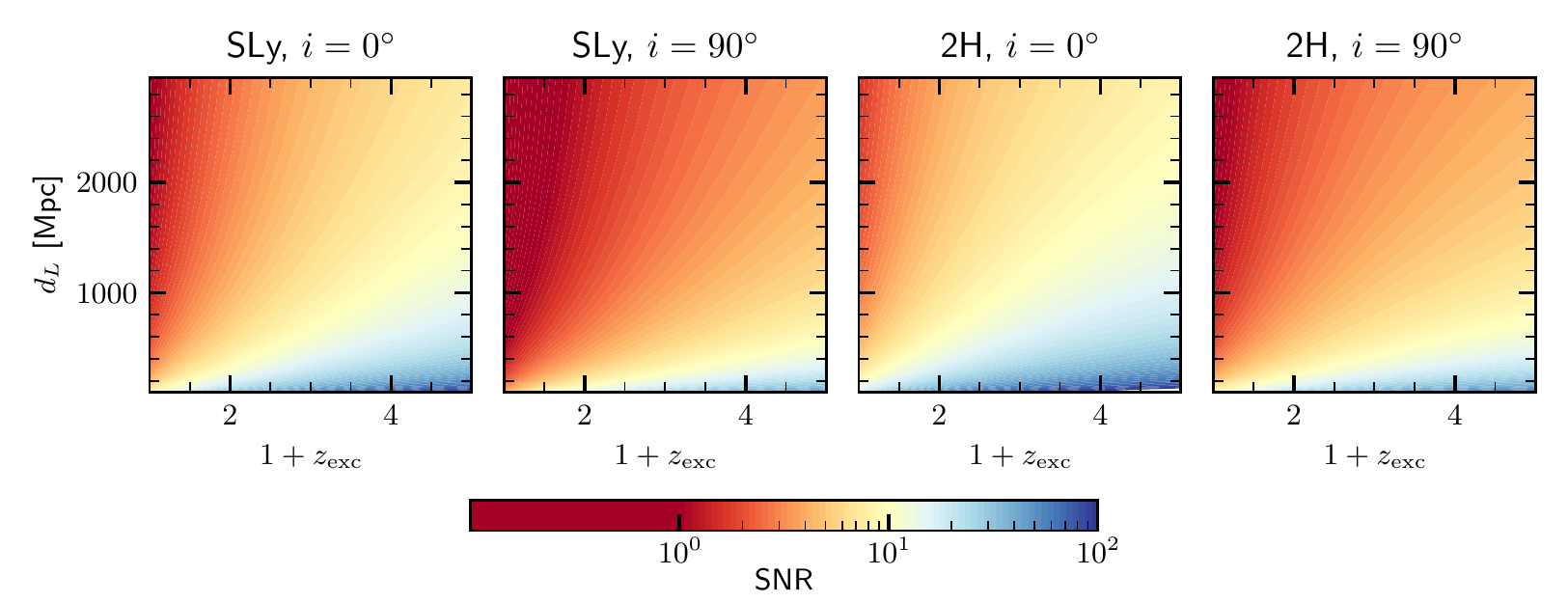}
	\caption{Postmerger SNRs for face-on and edge-on binaries for the Advanced LIGO (top row) and Cosmic Explorer (bottom row) sensitivities.
	As expected, SNR increases with an increase in the $ \onepluszextra $ and a decrease in the luminosity distance $ d_L $.
	For aLIGO sensitivity and $ \onepluszextra=4 $, typical SNRs are $ \sim 2 \ (0.2) $ for face-on (edge-on) binaries.
	However, for the Cosmic Explorer sensitivity, the typical SNRs for the same value of $ \onepluszextra $ are $\sim 10 \ (3)$.
	We also note that in both cases, the SNR for the ``no-redshifting'' case is always $< 1$.
	}
	\label{fig:SNR_calculation}
	
\end{figure*}

We use the methods described in Section \ref{sec:method} to quantify improvements gained by redshifting of the GW signal.
We first calculate the SNR of the postmerger signal using the prescription mentioned in Section \ref{subsec:postmerger-snr}. 
We use the numerical relativity simulations BAM:0002 and BAM:0098 \citep{Bernuzzi:2014kca} from the Computational Relativity (CoRe) database \citep{CoRe}. These assume EOSs 2H \citep{2H} and SLy \citep{SLy} respectively \footnote{2H and SLy are chosen to be representative of stiff and soft EOSs, respectively.}, and model the last few orbits of the binary neutron star inspiral, merger, and postmerger phases.  The component masses are set to $1.35 M_\odot$ and $1.35 M_\odot$, and the effective tidal deformabilities are $ \kappa^T_2 = 436$ and $ \kappa^T_2 = 73$.

We consider two observing scenarios \citep{observer_summary, Voyager_PSD, ET_PSD, CE_PSD}: 
\begin{enumerate}
\item {\it O5}: The fifth observing run of the LIGO-Virgo-Kagra network, consisting of 3 LIGO detectors (including LIGO-India \citep{LIGO-INDIA, Saleem:2021iwi}), and the Virgo \citep{advvirgo} and KAGRA \citep{KAGRA:2020tym} detectors, operating at their A+ sensitivities.
\item {\it 3G}: The third generation network, consisting of two Cosmic Explorers \citep{CE}  located at Hanford and Livingston, as well as one Einstein Telescope \citep{ET} located at Virgo's site. 
\end{enumerate}
We then calculate the optimal SNR for the postmerger on a grid of excess redshifts $ \onepluszextra$ and $ d_L $.  

Figure \ref{fig:SNR_calculation} shows the results from the SNR calculation for the O5 (top row) and 3G (bottom row) scenarios.  As expected, we see that the SNR of the postmerger increases with an increase in $\onepluszextra$ and decreases with an increase in $ d_L $. 
For SLy, when the binary is face-on \textit{ie.} inclination  $ i = 0 \degree $, the SNRs are all close to 1 or below when there is no excess redshifting, for systems located at $150$ Mpc or larger. However, increasing $ \onepluszextra $ to a value of 2.5 increases the SNRs to values well above 1, with some being as high as 6 at $200$ Mpc. In the case of edge-on binary $ (i = 90 \degree) $, without excess redshifting, all binaries at distances $> 100$ Mpc have postmergers with SNRs below 1. However, a $\onepluszextra=4$ would provide an SNR of $\sim 5$ at $100$ Mpc. 

The situation is more optimistic for the 2H EOS, with  $\onepluszextra=2.5$ providing an SNR of $\sim 10$ even at $200$ Mpc for the face-on binary. The edge-on case has lower SNRs as expected, although larger than such systems with an SLy EOS. Nevertheless, given that detecting the postmerger in O5 requires large excess redshift values, even for distances within $500$ Mpc, it seems unrealistic that the postmerger will be detected in O5,and if detected, will likely occur due to the BNS merger being within $\mathcal{O}(10)$ Mpc.

The impact of excess redshifting is more drastic in the 3G scenario.  Even $\onepluszextra = 2.5$ enables a detection with SNRs of $\approx 10$ and $\approx 20$ at $d_L = 1000$ Mpc, for the SLy and 2H EOSs respectively, assuming face-on binaries. Edge-on systems can also be detected with SNRs of $\sim 5$ and $\sim 8$ for the two EOSs respectively. 

\begin{figure*}
	
	\includegraphics[width=\textwidth]{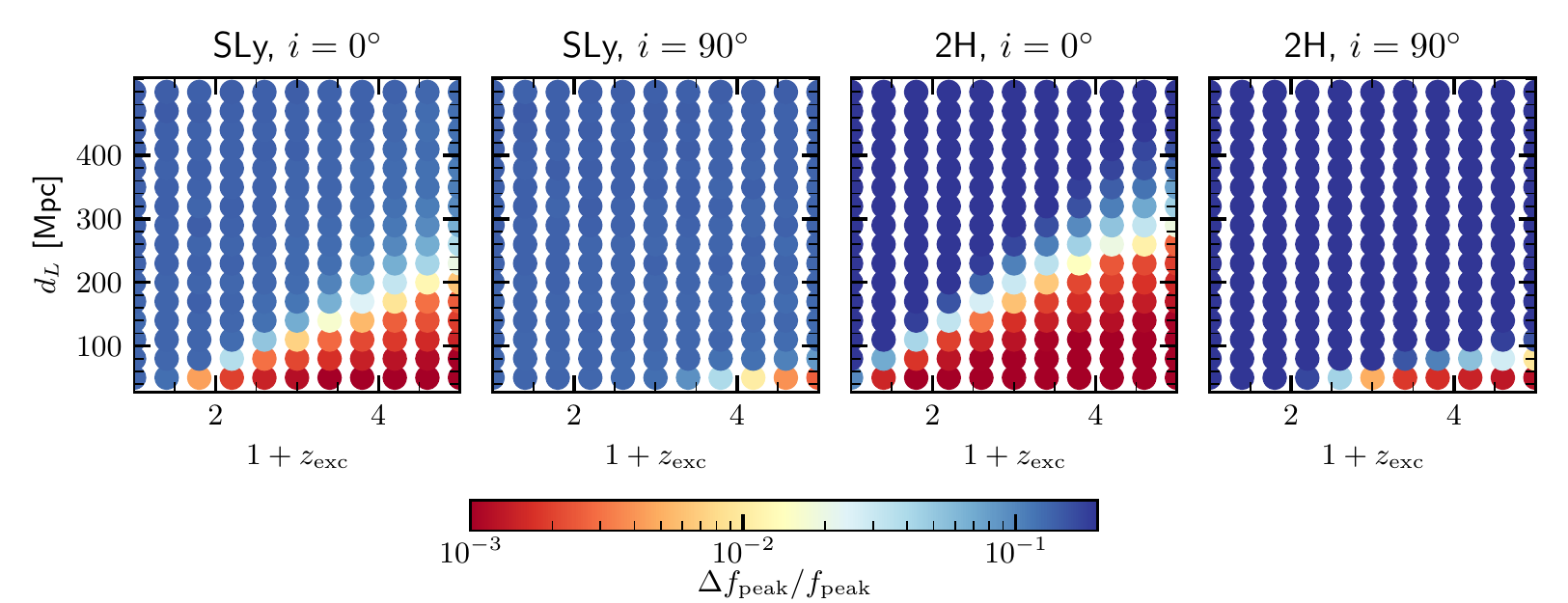}
	\includegraphics[width=\textwidth]{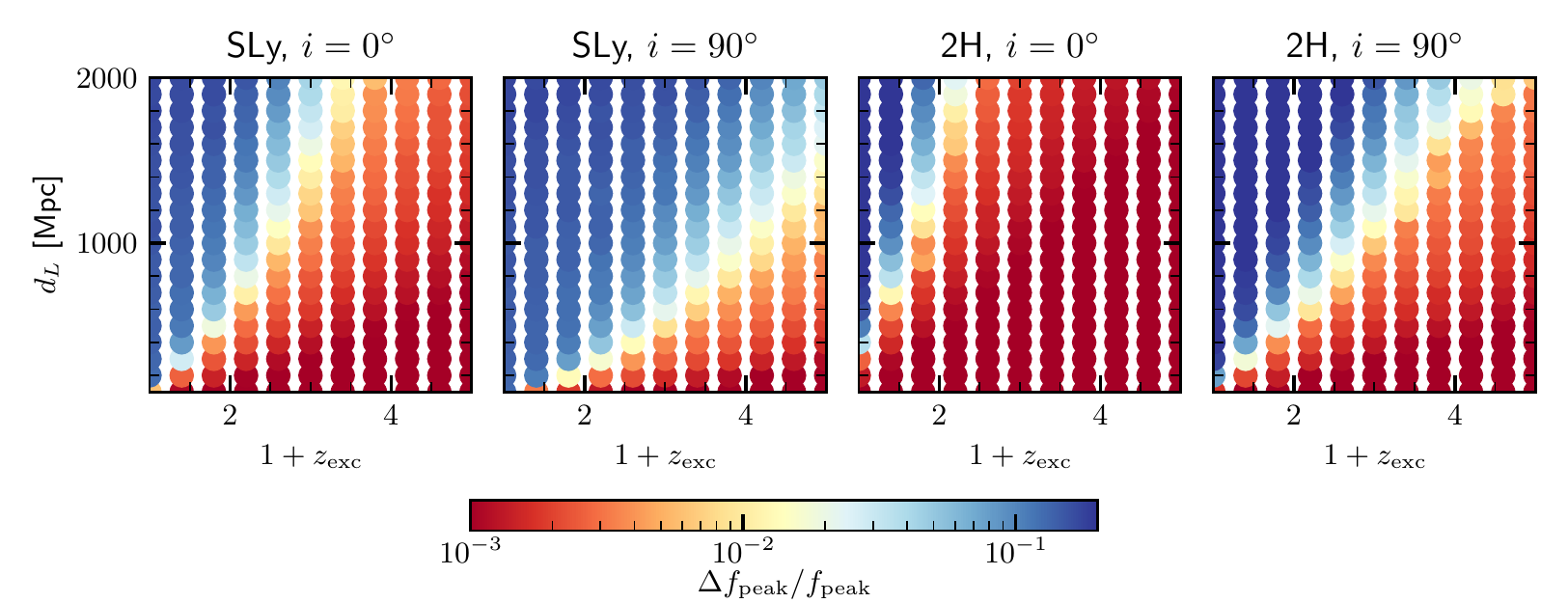}
	\caption{Relative errors in the measurement of the peak frequency for O5 (Top Panel) and 3G (Bottom Panel), for a range of excess redshifts and luminosity distances. The relative errors reduce with increasing excess redshift, decreasing luminosity distance, and stiffer EOS. In O5, the relative error for all configurations considered is $\sim 10^{-1}$ without excess redshift, even for binaries as close as $40$ Mpc, suggesting that the sampling of the posterior on model parameters (in particular, $f_\mathrm{peak}$) has returned the prior. On the other hand, even a mild excess redshift of $1.5$ can constrain the peak to relative errors of $\sim 10^{-2}$ at $40$ Mpc.  In 3G, an excess redshift of $\onepluszextra = 2.5$ could constrain the peak frequency for the SLy EOS to $\sim 10^{-2}$, at $1000$ Mpc, assuming face-on orbits. This constraint is tighter by an order of magnitude for the 2H EOS.  }
	\label{fig:errors_in_peak_frequency}
	
\end{figure*}

Having established the impact of excess redshifting on the detectability of the postmerger signal, we now investigate how this will improve the measurement of parameters related to the postmerger, in particular, the postmerger peak frequency.

We inject the phenomenological model \citep{Tsang2019} outlined in Section \ref{subsec:pheno-model} into a ``zero-noise'' realization for the detectors, again on a grid of $ \onepluszextra $ and $ d_L $. 
The ``zero-noise'' realization can be thought of as the most probable realization of the noise, and the error estimates obtained from this realization will be similar to those obtained using non-zero realizations. The model is fitted to the BAM:0002 and BAM:0098 postmerger peaks, and the fits are then used as proxys for the 2H and SLy postmerger waveforms. 

For each injection on the grid, we perform Bayesian parameter estimation to find the recovered posteriors and relative error on the peak frequency $ f_\mathrm{peak} $ and the spread in the frequency domain $ f_\mathrm{spread} $.
We fix the sky location and the extra redshift of the binary, thereby varying only $ f_\mathrm{peak} $ and $ f_\mathrm{spread} $ over a flat prior range between $[1800, 2200]$ and $[10, 30]$ respectively. We make use of the open source package \texttt{bilby} \citep{Ashton:2018jfp} coupled with the dynamical nested sampler \texttt{dynesty} \citep{Speagle:2019ivv} to streamline our parameter estimation analyses.

The results are summarized in Figure \ref{fig:errors_in_peak_frequency}, where the relative error in the measurement of the peak frequency is plotted as a function of excess redshift and luminosity distance. We see that in the O5 scenario, even for BNSs as close as $\sim 40$ Mpc, the postmerger peak frequency measurement has an error of $\sim 10^{-1}$ without redshifting \footnote{Note that errors of $10^{-1}$ imply that the sampling of the PE posterior has returned the prior.}.  On the other hand, even a mild $ \onepluszextra = 1.5$ gives a relative error of $\sim 10^{-2}$ at the same distance for the face-on BNS. Larger excess redshifts give similar relative errors out to larger luminosity distances.  As with the SNRs, the relative errors improve with increasing excess redshifts, decreasing luminosity distances, face-on orientation and stiffer EOS.

In the 3G scenario, while the general trends are similar to the O5 scenario, the relative errors are significantly smaller. In particular, $\onepluszextra \sim 2$ gives relative errors of $ < 10^{-1}$ even at $1000$ Mpc for the face-on SLy case, and $< 10^{-2}$ for the 2H EOS. The edge-on BNSs at the similar distances would require larger excess redshifts to acquire the same relative errors \footnote{Note that our parameter estimation exercise assumes that the degree of excess redshifting would be known by independent means (which we discuss in Sec.~\ref{sec:conclusion})}.  

\begin{figure*}
	
	\includegraphics{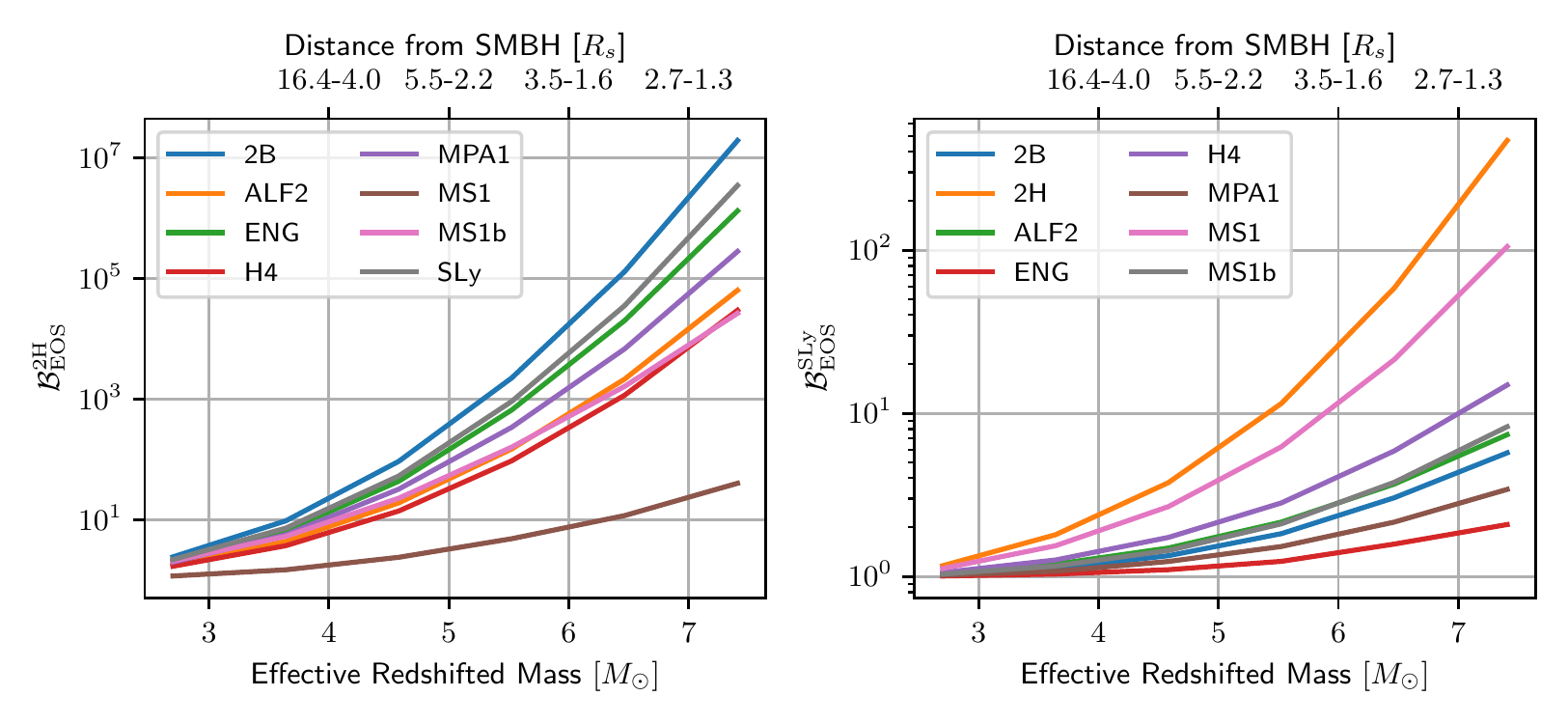}
	\includegraphics{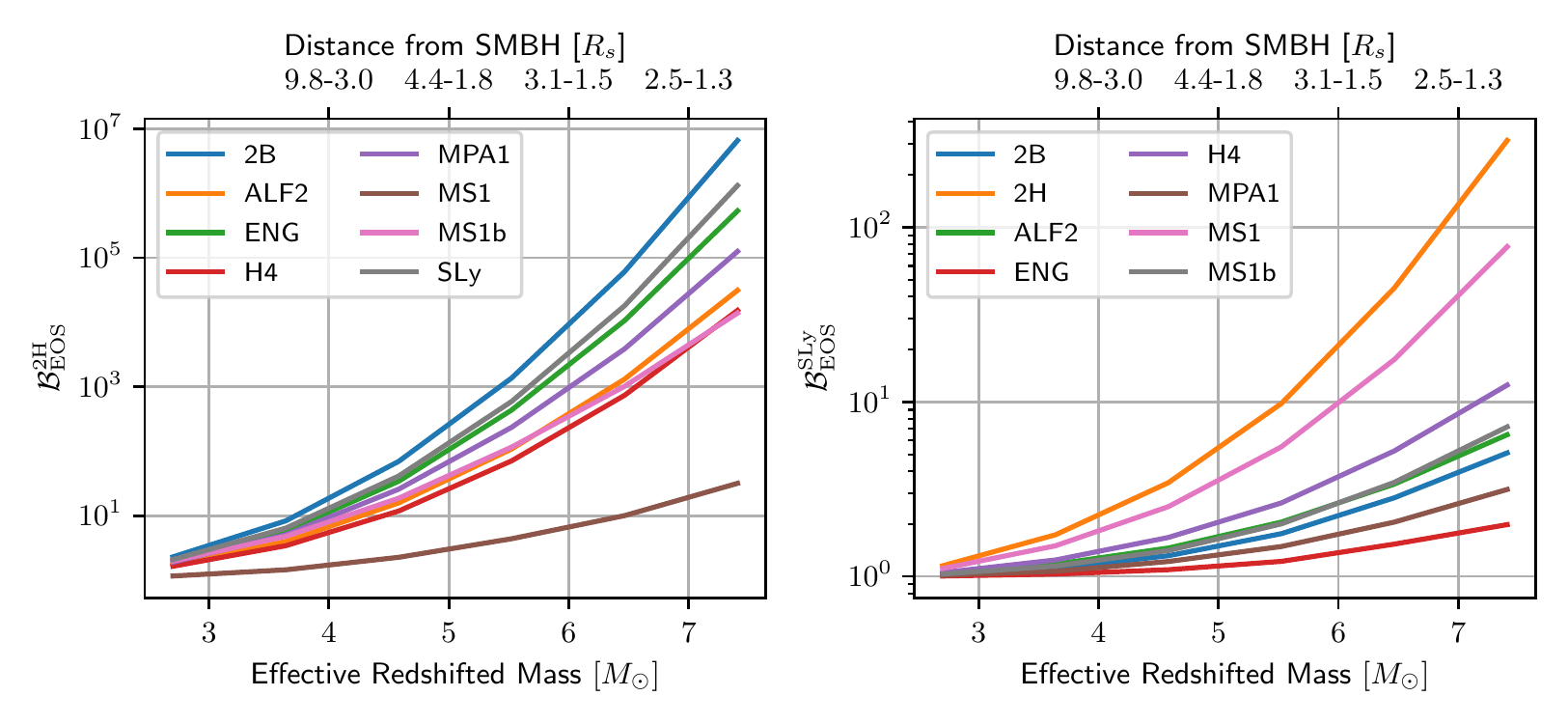}
	
	\caption{EOS model selection Bayes factors, as a function of effective redshifted mass (or, equivalently, the distance of the BNS from the SMBH.). The effective redshifted mass is simply $M(1 + z_{\mathrm{tot}})$. The Bayes factor $\mathcal{B}^{\mathrm{SLy}, 2\mathrm{H}}_{\mathrm{EOS}}$ assumes that SLy (2H) is the ``true'' (injected) EOS, and quantifies how well these EOSs can be distinguished from other EOS models.  The distance from the SMBH (on the top horizontal axis) pertains to the location of the BNS with respect to the SMBH at which the corresponding redshifted mass in the bottom horizontal axis is measured in the detector frame, assuming a circular geodesic motion in Kerr spacetime (with a spin of $a = 0.95$),  and standard cosmology.  This distance depends on the orientation of the outer binary with respect to the observer. The larger (smaller) value at each tick mark corresponds to purely longitudinal (transverse) motion of the BNS around the SMBH. The top (bottom) panels show the results for face-on (edge-on) BNSs located at $1000 (500)$ Mpc. An effective redshifted mass of $\sim 5 M_{\odot}$ improves the Bayes factor by a factor of $\sim 10 (> 100)$ for the SLy (2H) EOS, and face-on orbits. Such a detector frame mass is achieved at $\sim 6R_s$ or smaller, depending on the orientation of the outer binary. }
	\label{fig:bayes_factor}
	
\end{figure*}

\begin{table}
\centering
\caption{List of BNS numerical simulations used in this study, along with their EoSs,  effective tidal parameter in the EOB formalism $ \kappa^T_\mathrm{2} $, and the merger frequency calculated using Eq. \eqref{eqn:analytical-merger-frequency}. The simulations are part of the CoRe database, and were performed in  \citep{Bernuzzi:2014kca, Bernuzzi:2014owa, Dietrich:2017aum}}
\label{tab:EOS}
\begin{tabular}{llrr}
\toprule
 CoRe ID &  EOS &  $ \kappa^\mathrm{T}_{\mathrm{2}} $ &  $ f_\mathrm{merg} $ [Hz] \\
\midrule
BAM:0001 &   2B &                                  24 &                      2297 \\
BAM:0002 &   2H &                                 436 &                      1231 \\
BAM:0003 & ALF2 &                                 138 &                      1742 \\
BAM:0022 &  ENG &                                  90 &                      1922 \\
BAM:0035 &   H4 &                                 208 &                      1555 \\
BAM:0058 & MPA1 &                                 114 &                      1825 \\
BAM:0060 &  MS1 &                                 325 &                      1353 \\
BAM:0064 & MS1b &                                 287 &                      1408 \\
BAM:0098 &  SLy &                                  73 &                      2002 \\
\bottomrule
\end{tabular}
\end{table}

Finally, we quantify the improvement in the distinguishability of the EoS using the approximate Bayes factor described in Section \ref{subsec:eos-model-selection}  as our discriminator. 
We use numerical simulations of BNSs with masses $ m_1 = m_2 = 1.35 M_\odot $ and the same postmerger waveforms used above pertaining to the same two EOSs, as our fiducial ``true'' EOSs. We then evaluate Bayes factors ($\mathcal{B}^{\mathrm{SLy,2H}}_{\mathrm{EOS}}$) following the prescription delineated in Sec.~\ref{subsec:eos-model-selection} for a range of different EOSs.  The details of the numerical simulations used can be found in Table \ref{tab:EOS}. 

Figure \ref{fig:bayes_factor} shows the results from this investigation, for face-on and edge-on binaries, in the 3G era. The face-on binaries are assumed to be at $1000$ Mpc, and the edge-on binaries are assumed to be at $500$ Mpc. The horizontal axis plots the effective redshifted total mass, which is simply the source-frame mass multiplied by $1 + z_{\mathrm{tot}}$.  The vertical axis shows the Bayes factor between the ``true'' EOS and a host of other EOSs. The upper horizontal axis plots the distance from the SMBH at which the effective redshifted mass is acquired. The SMBH is assumed to have a spin of $a=0.95$, consistent with measured spins of SMBHs \citep{Reynolds2014}, and the motion is assumed to be a circular geodesic in the Kerr metric. The two numbers at each x-value correspond to the minimum and maximum distance from the SMBH at which the same total redshift occurs -- the minimum corresponds to transverse motion devoid of Doppler redshifting, and the maximum corresponds to longitudinal motion where the velocity vector of the orbiting binary points radially away from the Earth \footnote{Note that we only consider those BNSs whose velocities (pertaining to their motion around the SMBH) are assumed to have a component that points radially away from the Earth. There may of course be BNSs whose velocity component points towards the Earth, and the resulting distance from the SMBH may need to be even smaller than the minimum value computed here.}. 

With an effective redshifted total mass of $\sim 5 M_{\odot}$, we see that the SLy and 2H EOSs can be distinguished with $ \mathcal{B}^{\mathrm{2H}}_{{\mathrm{SLy}}} > 100~\mathrm{and}~\mathcal{B}^{\mathrm{SLy}}_{{\mathrm{2H}}} \gtrsim 10$ for $ d_L \leq 1000$ Mpc, assuming face-on orbits. \footnote{Note that even a small decrease in luminosity distance increases the Bayes factor significantly, since the log of the Bayes factor is inversely proportional to the square of the luminosity distance.}. At $d_L = 1000$ Mpc, an effective redshifted total mass of $\sim 5 M_{\odot}$ would need the merger to occur within $\sim 6 R_s$ from the SMBH, where $R_s$ is the Schwarzschild radius of the SMBH. 

The Bayes factors for edge-on binaries at $500$ Mpc are similar to those of face-on binaries at $1000$ Mpc, although the smaller luminosity distance reduces the cosmological redshift. Thus,  larger excess redshifts are required to produce the same effective redshifted mass. This, in turn, requires mergers to occur relatively closer to the SMBH, as compared to mergers at $1000$ Mpc.

Clearly, then, even in the 3G era, EOS model selection using postmerger signals out to luminosity distances of $\sim 1 \mathrm{Gpc}$ requires excess-redshifting, without which, Bayes factors are found to be $\sim 1$.

\section{Discussion}\label{sec:conclusion}
	\subsection{Executive summary}
Extracting astrophysical information about NS properties from GWs emanated during the postmerger phase of BNS coalescences could prove to be crucial to our understanding of NSs and the nature of matter at extreme densities. The high-frequency noise in current GW detectors makes the detection of such signals almost impossible, except perhaps for unrealistically nearby BNSs. Even 3G detectors could struggle to constrain properties of the postmerger---such as its peak frequency---for BNSs located at distances greater than a few tens of Mpc.

In this work, we show that, in the absence of a detector sensitive at $\mathcal{O}(1000)$ Hz, one way to detect the BNS postmerger and constrain its properties, for BNSs located at distances of several tens to hundreds of Mpc, is for the postmerger to be somehow ``dragged'' into the band of the detectors.  Perhaps the only conceivable astrophysical mechanism to achieve this is for the BNS to merge in the vicinity of an SMBH.  The gravitational redshift produced by the gravitational field of the SMBH, as well as the Doppler redshift due to the BNS's proper motion around the SMBH (assuming that its velocity has a component pointing radially away from the Earth), in addition to the cosmological redshift, could effectively stretch the postmerger GW towards frequencies where ground based detectors are sensitive. 

Denoting the combined gravitational and Doppler redshifts as $\onepluszextra$, we vary this quantity from $1-5$ to see its effect on the postmerger SNR, the width of the Bayesian posterior on the peak frequency, as well as EOS model selection. We use NR simulations of postmerger signals for non-spinning $1.35-1.35 M_{\odot}$ BNSs with the SLy and 2H EOSs to evaluate the optimal SNR for a range of luminosity distances.  Using a phenomenological fit to the postmerger waveforms, we evaluate a posterior on the postmerger peak frequency. We then estimate approximate Bayes factors to discriminate between SLy (2H) and other EOSs. We perform these exercises for varying excess redshifts, luminosity distances and both EOSs, as well as two observing scenarios (O5 and 3G).

We find that the SNRs, relative errors on the peak frequency, and Bayes factors, improve significantly with increasing excess redshift. In O5, postmergers could optimistically be detectable out to $150$ Mpc with $\onepluszextra=2.5$, and the peak frequency constrained to well within a relative error of $10^{-1}$, which would not be the case without this extra redshift. In 3G, similar values are found for distances up to $1000$ Mpc. In addition, EOS model selection (specifically, the Bayes factor) improves by up to an order of magnitude, at 1000 Mpc, assuming the ``true'' EOS to be SLy.  For a stiffer (2H) EOS, this improvement factor jumps to more than two orders of magnitude.

\subsection{Measuring the excess redshift}
While measuring the postmerger SNR, or evaluating the EOS model selection Bayes factors \footnote{In this work, the evaluation of the Bayes factors assumes that there is no degeneracy between the total mass of the BNS, and the EOS. In other words, given a postmerger signal pertaining to an EOS and a total mass, no other EOS can exactly replicate all the spectral features of this signal for {\it any} total mass.}, do not require an estimate of $\onepluszextra$, estimating the true value of the postmerger peak requires it, to break the degeneracy between $f_\mathrm{peak}$ and $1 + z_{\mathrm{tot}}$. 

There are at least three methods that could potentially disentangle these two quantities.  Perhaps the most comprehensive method would involve a multiband detection. Information about the outer binary's orbit around the SMBH would be acquired by the space-based low-frequency detector LISA \citep{Danzmann1996}, while ground based 3G detectors would probe the motion of the inner binary. If, in addition, the BNS merger's counterpart was observed, then multiple measurements of the mass of the BNS, as well as the degree of redshifting, would allow for consistency checks and a robust estimate of the true location of $f_\mathrm{peak}$.

The second method uses the measurement of tidal parameters from the inspiral to break the degeneracy between masses and redshift. This can be done using quasi-universal relations between the source frame mass and the tidal parameters \citep{Chatterjee:2021xrm}, or combining information of the inspiral with the measurement of the postmerger \citep{Messenger2014}. However, it must be noted that while these methods are relatively straightforward to employ for an isolated merger, mergers in the vicinity of the SMBH would need to account for the imprint of the BNS's proper motion around the SMBH on the inspiral waveform. This is especially true for the 3G scenario, where the inspiral in-band could last for $\mathcal{O}(10-100)$ minutes.

Yet another method would be to use the BNS merger's counterpart (kilonovae, short gamma-ray bursts) \citep{LIGOScientific:2017ync, Zhu:2020wrd} assisted by a galaxy catalog to identify the host galaxy, thereby estimating $ z_\mathrm{cos} $. Comparing the associated $ d_L $ with the measurement of $ d_L $ from the merger's GWs would enable an estimate of $ \onepluszextra{} $ (cf. Eq.~\ref{eq:redshift_relation}).  The ``unredshifted'' $ d_L $ could also be estimated using the kilonova as a standard candle \citep{Kashyap2019, Coughlin2020}, although such a method would need to be K-corrected \citep{Kim:1995qj}. Note, however, that these estimates of $d_L$ would have associated error-bars, which would need to be accounted for when converting from the redshifted postmerger peak to its true value. Thus, the relative errors on $f_\mathrm{peak}$ quoted in this work can be thought of as lower-limits.

\subsection{Outlook}
We showed in this work that {\it if} a BNS merger occurred in the vicinity of an SMBH, then the redshifted postmerger could not only be detected, but its spectral properties could potentially be constrained, and the NS EOS could be determined.  While we speculated (based on recent simulation results) in the Sec.~\ref{sec:introduction} that detecting BNS postmergers redshifted by an SMBH might not be inconceivable (especially in the 3G era),  the rate of such systems is far from determined. Achieving this would require a full understanding of competing mechanisms in dense stellar environments. While some such as tidal-capture, tidal-perturbation, and migration traps close to the ISCO, could enable BNS mergers near SMBHs, others such as mass-segregation and kick-velocities inhibit the occurrence of such events.  Nevertheless, an approximate (and likely optimistic) upper limit can be estimated as follows. 

With tidal capture, the rate of stellar mass BBH mergers within $10 R_s$ of an SMBH is estimated to be $\sim 0.03 \mathrm{Gpc}^{-3}\mathrm{yr}^{-1}$ \citep{ChenLiCao2019}.  On the other hand, rate of mergers within the last migration trap (which can be as close as $6 R_s$ from a non-spinning SMBH) is estimated to be $\sim 0.4 \mathrm{Gpc}^{-3}\mathrm{yr}^{-1}$ \citep{Peng2021}. However, the rate of BNS mergers within $10 R_s$ has (to the best of our knowledge) not been computed. An approximate upper limit on this rate can be arrived at using the results of \citet{McKernan2020}, where the upper limit on the ratio of the number of mergers of BNSs to BBHs within AGNs is estimated to be $\sim 4$. Assuming that this fraction is also maintained in the vicinity of the SMBH, the rate of BNS mergers would be $ \lesssim 2 \mathrm{Gpc}^{-3}\mathrm{yr}^{-1}$.

From Fig.~\ref{fig:errors_in_peak_frequency}, assuming the more realistc SLy EOS, constraining the postmerger peak frequency would be difficult for non-excess-redshifted BNSs located at distances larger than $\sim 100$ Mpc in the 3G era. 
Thus, using the limits on the rate of BNS mergers acquired from LIGO-Virgo data ($10-1700 \mathrm{Gpc}^{-3}\mathrm{yr}^{-1}$ \citep{gwtc3-rates}), the number of BNSs (per year) that enable such constraints could be anywhere between about $0-7$. On the other hand, the peak frequency could be well constrained for BNSs out to distances of $\sim 1000$ Mpc if they're excess-redshifted. The upper limit on their merger rate, as estimated above, suggests that within $1000$ Mpc, $0-2$ such events per year could occur. This is a non-trivial fraction of the number of non-redshifted BNS mergers per year within $100$ Mpc.

It must however be noted that not {\it all} BNSs in the vicinity of SMBHs will necessarily be sufficiently excess-redshifted to present a detectable waveform with constrainable spectral properties. In fact, a fraction may even be blue-shifted. Nevertheless, even if $\mathcal{O}(1)$ such events were detected by the end of the 10-year 3G era, they could provide unprecedented access to postmergers and their astrophysics.

In principle, the non-detection of redshifted BNS postmergers could provide upper limits on the rate of such events. While admittedly, these rates would be limited to specific masses and EOSs, for which numerical relativity based postmerger waveforms are available, they could enable constraints that could guide models of mergers in the neighbourhood of SMBHs. We are currently working on estimating rate upper limits using publicly available LIGO-Virgo data from its first three observing runs. We hope to report these results soon.

\section*{Acknowledgements}
We thank Simon Stevenson for a very careful reading of our manuscript. We also thank Sukanta Bose, Bala Iyer, Nathan Johnson-McDaniel, R. Loganayagam,  Sanjit Mitra, Suvrat Raju and Shiv Sethi for many useful discussions, and members of the gravitational-wave groups at ICTS and IUCAA for feedback. Computations were performed on the ICTS Alice Cluster. This work makes use of \texttt{NumPy} \citep{vanderWalt:2011bqk}, \texttt{SciPy} \citep{Virtanen:2019joe}, \texttt{astropy} \citep{2013A&A...558A..33A, 2018AJ....156..123A}, \texttt{Matplotlib} \citep{Hunter:2007}, \texttt{jupyter} \citep{jupyter}, \texttt{pandas} \citep{mckinney-proc-scipy-2010} \texttt{dynesty} \citep{Speagle:2019ivv}, \texttt{bilby} \citep{Ashton:2018jfp} and \texttt{PESummary} \citep{Hoy:2020vys} software packages.
S. J. K.’s work was supported by a grant from the Simons Foundation (Grant No. 677895, R. G.). P. A.’s research was supported by the Max Planck Society through a Max Planck Partner Group at ICTS and by the Canadian Institute for Advanced Research through the CIFAR Azrieli Global Scholars program. This work is also supported by the Department of Atomic Energy, Government of India, under Project No. RTI4001. 

\section*{Data Availability}
The data underlying this article will be shared on reasonable request to the corresponding author.

\bibliography{references}

\begin{thebibliography}{}
\makeatletter
\relax
\def\mn@urlcharsother{\let\do\@makeother \do\$\do\&\do\#\do\^\do\_\do\%\do\~}
\def\mn@doi{\begingroup\mn@urlcharsother \@ifnextchar [ {\mn@doi@}
  {\mn@doi@[]}}
\def\mn@doi@[#1]#2{\def\@tempa{#1}\ifx\@tempa\@empty \href
  {http://dx.doi.org/#2} {doi:#2}\else \href {http://dx.doi.org/#2} {#1}\fi
  \endgroup}
\def\mn@eprint#1#2{\mn@eprint@#1:#2::\@nil}
\def\mn@eprint@arXiv#1{\href {http://arxiv.org/abs/#1} {{\tt arXiv:#1}}}
\def\mn@eprint@dblp#1{\href {http://dblp.uni-trier.de/rec/bibtex/#1.xml}
  {dblp:#1}}
\def\mn@eprint@#1:#2:#3:#4\@nil{\def\@tempa {#1}\def\@tempb {#2}\def\@tempc
  {#3}\ifx \@tempc \@empty \let \@tempc \@tempb \let \@tempb \@tempa \fi \ifx
  \@tempb \@empty \def\@tempb {arXiv}\fi \@ifundefined
  {mn@eprint@\@tempb}{\@tempb:\@tempc}{\expandafter \expandafter \csname
  mn@eprint@\@tempb\endcsname \expandafter{\@tempc}}}

\bibitem[\protect\citeauthoryear{Aasi et~al.}{Aasi et~al.}{2015}]{advligo}
Aasi J.,  et~al., 2015, \mn@doi [Classical and Quantum Gravity]
  {10.1088/0264-9381/32/7/074001}, 32, 074001

\bibitem[\protect\citeauthoryear{Abbott et~al.}{Abbott et~al.}{2017a}]{CE_PSD}
Abbott B.~P.,  et~al., 2017a, \mn@doi [Classical and Quantum Gravity]
  {10.1088/1361-6382/aa51f4}, 34, 044001

\bibitem[\protect\citeauthoryear{Abbott et~al.}{Abbott
  et~al.}{2017b}]{GW170817-DETECTION}
Abbott B.~P.,  et~al., 2017b, \mn@doi [Phys. Rev. Lett.]
  {10.1103/PhysRevLett.119.161101}, 119, 161101

\bibitem[\protect\citeauthoryear{Abbott et~al.}{Abbott
  et~al.}{2017c}]{LIGOScientific:2017ync}
Abbott B.~P.,  et~al., 2017c, \mn@doi [Astrophys. J. Lett.]
  {10.3847/2041-8213/aa91c9}, 848, L12

\bibitem[\protect\citeauthoryear{{Abbott} et~al.,}{{Abbott}
  et~al.}{2017d}]{GW170817-PM}
{Abbott} B.~P.,  et~al., 2017d, \mn@doi [\apjl] {10.3847/2041-8213/aa9a35},
  \href {https://ui.adsabs.harvard.edu/abs/2017ApJ...851L..16A} {851, L16}

\bibitem[\protect\citeauthoryear{Abbott et~al.}{Abbott
  et~al.}{2018a}]{aplus_sensitivity}
Abbott B.~P.,  et~al., 2018a, \mn@doi [Living Rev. Rel.]
  {10.1007/s41114-020-00026-9}, 21, 3

\bibitem[\protect\citeauthoryear{{Abbott}, {Abbott}, {Abbott}, {Acernese},
  {LIGO Scientific Collaboration}  \& {Virgo Collaboration}}{{Abbott}
  et~al.}{2018b}]{GW170817-EOS}
{Abbott} B.~P.,  {Abbott} R.,  {Abbott} T.~D.,  {Acernese} F.,  {LIGO
  Scientific Collaboration}  {Virgo Collaboration} 2018b, \mn@doi [\prl]
  {10.1103/PhysRevLett.121.161101}, \href
  {https://ui.adsabs.harvard.edu/abs/2018PhRvL.121p1101A} {121, 161101}

\bibitem[\protect\citeauthoryear{Abbott et~al.}{Abbott et~al.}{2019}]{GWTC-1}
Abbott B.~P.,  et~al., 2019, \mn@doi [Physical Review X]
  {10.1103/PhysRevX.9.031040}, \href
  {https://ui.adsabs.harvard.edu/abs/2019PhRvX...9c1040A} {9, 031040}

\bibitem[\protect\citeauthoryear{Abbott et~al.}{Abbott
  et~al.}{2021a}]{LIGOScientific:2021djp}
Abbott R.,  et~al., 2021a, arXiv preprint arXiv:2111.03606

\bibitem[\protect\citeauthoryear{Abbott et~al.}{Abbott
  et~al.}{2021b}]{gwtc3-rates}
Abbott R.,  et~al., 2021b, arXiv preprint arXiv:2111.03634

\bibitem[\protect\citeauthoryear{{Abbott} et~al.,}{{Abbott}
  et~al.}{2021c}]{GWTC-2}
{Abbott} R.,  et~al., 2021c, \mn@doi [Physical Review X]
  {10.1103/PhysRevX.11.021053}, \href
  {https://ui.adsabs.harvard.edu/abs/2021PhRvX..11b1053A} {11, 021053}

\bibitem[\protect\citeauthoryear{Acernese et~al.}{Acernese
  et~al.}{2014}]{advvirgo}
Acernese F.,  et~al., 2014, \mn@doi [Classical and Quantum Gravity]
  {10.1088/0264-9381/32/2/024001}, 32, 024001

\bibitem[\protect\citeauthoryear{{Ackley} et~al.,}{{Ackley}
  et~al.}{2020}]{NEMO}
{Ackley} K.,  et~al., 2020, \mn@doi [\pasa] {10.1017/pasa.2020.39}, \href
  {https://ui.adsabs.harvard.edu/abs/2020PASA...37...47A} {37, e047}

\bibitem[\protect\citeauthoryear{Aghanim et~al.,}{Aghanim
  et~al.}{2020}]{Planck2018}
Aghanim N.,  et~al., 2020, Astronomy \& Astrophysics, 641, A6

\bibitem[\protect\citeauthoryear{Akutsu et~al.}{Akutsu
  et~al.}{2021}]{KAGRA:2020tym}
Akutsu T.,  et~al., 2021, \mn@doi [PTEP] {10.1093/ptep/ptaa125}, 2021, 05A101

\bibitem[\protect\citeauthoryear{{Antonini} \& {Perets}}{{Antonini} \&
  {Perets}}{2012}]{Antonini2012}
{Antonini} F.,  {Perets} H.~B.,  2012, \mn@doi [\apj]
  {10.1088/0004-637X/757/1/27}, \href
  {https://ui.adsabs.harvard.edu/abs/2012ApJ...757...27A} {757, 27}

\bibitem[\protect\citeauthoryear{{Artymowicz}, {Lin}  \&
  {Wampler}}{{Artymowicz} et~al.}{1993}]{Artymowicz1993}
{Artymowicz} P.,  {Lin} D.~N.~C.,   {Wampler} E.~J.,  1993, \mn@doi [\apj]
  {10.1086/172690}, \href
  {https://ui.adsabs.harvard.edu/abs/1993ApJ...409..592A} {409, 592}

\bibitem[\protect\citeauthoryear{Ashton et~al.}{Ashton
  et~al.}{2019}]{Ashton:2018jfp}
Ashton G.,  et~al., 2019, \mn@doi [Astrophys. J. Suppl.]
  {10.3847/1538-4365/ab06fc}, 241, 27

\bibitem[\protect\citeauthoryear{{Astropy Collaboration} et~al.,}{{Astropy
  Collaboration} et~al.}{2013}]{2013A&A...558A..33A}
{Astropy Collaboration} et~al., 2013, \mn@doi [\aap]
  {10.1051/0004-6361/201322068}, \href
  {https://ui.adsabs.harvard.edu/abs/2013A&A...558A..33A} {558, A33}

\bibitem[\protect\citeauthoryear{{Astropy Collaboration} et~al.,}{{Astropy
  Collaboration} et~al.}{2018}]{2018AJ....156..123A}
{Astropy Collaboration} et~al., 2018, \mn@doi [\aj] {10.3847/1538-3881/aabc4f},
  \href {https://ui.adsabs.harvard.edu/abs/2018AJ....156..123A} {156, 123}

\bibitem[\protect\citeauthoryear{Bernuzzi, Nagar, Balmelli, Dietrich  \&
  Ujevic}{Bernuzzi et~al.}{2014}]{Bernuzzi:2014kca}
Bernuzzi S.,  Nagar A.,  Balmelli S.,  Dietrich T.,   Ujevic M.,  2014, \mn@doi
  [Phys. Rev. Lett.] {10.1103/PhysRevLett.112.201101}, 112, 201101

\bibitem[\protect\citeauthoryear{Bernuzzi, Nagar, Dietrich  \& Damour}{Bernuzzi
  et~al.}{2015}]{Bernuzzi:2014owa}
Bernuzzi S.,  Nagar A.,  Dietrich T.,   Damour T.,  2015, \mn@doi [Phys. Rev.
  Lett.] {10.1103/PhysRevLett.114.161103}, 114, 161103

\bibitem[\protect\citeauthoryear{Breschi, Bernuzzi, Zappa, Agathos, Perego,
  Radice  \& Nagar}{Breschi et~al.}{2019}]{Breschi:2019srl}
Breschi M.,  Bernuzzi S.,  Zappa F.,  Agathos M.,  Perego A.,  Radice D.,
  Nagar A.,  2019, \mn@doi [Phys. Rev. D] {10.1103/PhysRevD.100.104029}, 100,
  104029

\bibitem[\protect\citeauthoryear{Chatterjee, R., Holder, Holz, Perkins, Yagi
  \& Yunes}{Chatterjee et~al.}{2021}]{Chatterjee:2021xrm}
Chatterjee D.,  R. A. H.~K.,  Holder G.,  Holz D.~E.,  Perkins S.,  Yagi K.,
  Yunes N.,  2021, \mn@doi [Phys. Rev. D] {10.1103/PhysRevD.104.083528}, 104,
  083528

\bibitem[\protect\citeauthoryear{{Chatziioannou}}{{Chatziioannou}}{2020}]{Chatziioannou2020}
{Chatziioannou} K.,  2020, \mn@doi [General Relativity and Gravitation]
  {10.1007/s10714-020-02754-3}, \href
  {https://ui.adsabs.harvard.edu/abs/2020GReGr..52..109C} {52, 109}

\bibitem[\protect\citeauthoryear{Chen \& Han}{Chen \& Han}{2018}]{ChenHan2018}
Chen X.,  Han W.-B.,  2018, Communications Physics, 1, 1

\bibitem[\protect\citeauthoryear{{Chen}, {Li}  \& {Cao}}{{Chen}
  et~al.}{2019}]{ChenLiCao2019}
{Chen} X.,  {Li} S.,   {Cao} Z.,  2019, \mn@doi [\mnras]
  {10.1093/mnrasl/slz046}, \href
  {https://ui.adsabs.harvard.edu/abs/2019MNRAS.485L.141C} {485, L141}

\bibitem[\protect\citeauthoryear{{Clark}, {Bauswein}, {Stergioulas}  \&
  {Shoemaker}}{{Clark} et~al.}{2016}]{Clark2016}
{Clark} J.~A.,  {Bauswein} A.,  {Stergioulas} N.,   {Shoemaker} D.,  2016,
  \mn@doi [Classical and Quantum Gravity] {10.1088/0264-9381/33/8/085003},
  \href {https://ui.adsabs.harvard.edu/abs/2016CQGra..33h5003C} {33, 085003}

\bibitem[\protect\citeauthoryear{{CoRe Collaboration}}{{CoRe
  Collaboration}}{2022}]{CoRe}
{CoRe Collaboration} 2022, Computational Relativity ({CoRe}) GW database,
  \url{http://www.computational-relativity.org/gwdb/}

\bibitem[\protect\citeauthoryear{Cornish, Sampson, Yunes  \& Pretorius}{Cornish
  et~al.}{2011}]{Cornish2011}
Cornish N.,  Sampson L.,  Yunes N.,   Pretorius F.,  2011, Physical Review D,
  84, 062003

\bibitem[\protect\citeauthoryear{{Coughlin} et~al.,}{{Coughlin}
  et~al.}{2020}]{Coughlin2020}
{Coughlin} M.~W.,  et~al., 2020, \mn@doi [Physical Review Research]
  {10.1103/PhysRevResearch.2.022006}, \href
  {https://ui.adsabs.harvard.edu/abs/2020PhRvR...2b2006C} {2, 022006}

\bibitem[\protect\citeauthoryear{Creighton \& Anderson}{Creighton \&
  Anderson}{2012}]{creighton2012}
Creighton J.~D.,  Anderson W.~G.,  2012, Gravitational-wave physics and
  astronomy: An introduction to theory, experiment and data analysis.
John Wiley \& Sons

\bibitem[\protect\citeauthoryear{Cutler \& Flanagan}{Cutler \&
  Flanagan}{1994}]{cutler1994}
Cutler C.,  Flanagan E.~E.,  1994, Physical Review D, 49, 2658

\bibitem[\protect\citeauthoryear{Danzmann, Team  et~al.}{Danzmann
  et~al.}{1996}]{Danzmann1996}
Danzmann K.,  Team L.~S.,   et~al., 1996, Classical and Quantum Gravity, 13,
  A247

\bibitem[\protect\citeauthoryear{Dietrich, Bernuzzi  \& Tichy}{Dietrich
  et~al.}{2017}]{Dietrich:2017aum}
Dietrich T.,  Bernuzzi S.,   Tichy W.,  2017, \mn@doi [Phys. Rev. D]
  {10.1103/PhysRevD.96.121501}, 96, 121501

\bibitem[\protect\citeauthoryear{{Douchin} \& {Haensel}}{{Douchin} \&
  {Haensel}}{2001}]{SLy}
{Douchin} F.,  {Haensel} P.,  2001, \mn@doi [\aap]
  {10.1051/0004-6361:20011402}, \href
  {https://ui.adsabs.harvard.edu/abs/2001A&A...380..151D} {380, 151}

\bibitem[\protect\citeauthoryear{Easter, Lasky, Casey, Rezzolla  \&
  Takami}{Easter et~al.}{2019}]{Easter:2018pqy}
Easter P.~J.,  Lasky P.~D.,  Casey A.~R.,  Rezzolla L.,   Takami K.,  2019,
  \mn@doi [Phys. Rev. D] {10.1103/PhysRevD.100.043005}, 100, 043005

\bibitem[\protect\citeauthoryear{Flanagan \& Hinderer}{Flanagan \&
  Hinderer}{2008}]{Flanagan:2007ix}
Flanagan E.~E.,  Hinderer T.,  2008, \mn@doi [Phys. Rev. D]
  {10.1103/PhysRevD.77.021502}, 77, 021502

\bibitem[\protect\citeauthoryear{Ford \& McKernan}{Ford \&
  McKernan}{2021}]{Ford2021}
Ford K.,  McKernan B.,  2021, arXiv preprint arXiv:2109.03212

\bibitem[\protect\citeauthoryear{{Freitag}, {Amaro-Seoane}  \&
  {Kalogera}}{{Freitag} et~al.}{2006}]{Freitag2006}
{Freitag} M.,  {Amaro-Seoane} P.,   {Kalogera} V.,  2006, \mn@doi [\apj]
  {10.1086/506193}, \href
  {https://ui.adsabs.harvard.edu/abs/2006ApJ...649...91F} {649, 91}

\bibitem[\protect\citeauthoryear{{Gendreau} et~al.,}{{Gendreau}
  et~al.}{2016}]{NICERa}
{Gendreau} K.~C.,  et~al., 2016, in {den Herder} J.-W.~A.,  {Takahashi} T.,
  {Bautz} M.,  eds,  Society of Photo-Optical Instrumentation Engineers (SPIE)
  Conference Series Vol. 9905, Space Telescopes and Instrumentation 2016:
  Ultraviolet to Gamma Ray. p. 99051H, \mn@doi{10.1117/12.2231304}

\bibitem[\protect\citeauthoryear{{Goodman} \& {Tan}}{{Goodman} \&
  {Tan}}{2004}]{Goodman2004}
{Goodman} J.,  {Tan} J.~C.,  2004, \mn@doi [\apj] {10.1086/386360}, \href
  {https://ui.adsabs.harvard.edu/abs/2004ApJ...608..108G} {608, 108}

\bibitem[\protect\citeauthoryear{Hild}{Hild}{2012}]{ET_PSD}
Hild S.,  2012, \mn@doi [Classical and Quantum Gravity]
  {10.1088/0264-9381/29/12/124006}, 29, 124006

\bibitem[\protect\citeauthoryear{{Holz} \& {Hughes}}{{Holz} \&
  {Hughes}}{2005}]{HolzHughes2005}
{Holz} D.~E.,  {Hughes} S.~A.,  2005, \mn@doi [\apj] {10.1086/431341}, \href
  {https://ui.adsabs.harvard.edu/abs/2005ApJ...629...15H} {629, 15}

\bibitem[\protect\citeauthoryear{{Hotokezaka}, {Kiuchi}, {Kyutoku},
  {Muranushi}, {Sekiguchi}, {Shibata}  \& {Taniguchi}}{{Hotokezaka}
  et~al.}{2013}]{HotokezakaHNS}
{Hotokezaka} K.,  {Kiuchi} K.,  {Kyutoku} K.,  {Muranushi} T.,  {Sekiguchi}
  Y.-i.,  {Shibata} M.,   {Taniguchi} K.,  2013, \mn@doi [\prd]
  {10.1103/PhysRevD.88.044026}, \href
  {https://ui.adsabs.harvard.edu/abs/2013PhRvD..88d4026H} {88, 044026}

\bibitem[\protect\citeauthoryear{{Hoy} \& {Raymond}}{{Hoy} \&
  {Raymond}}{2021}]{Hoy:2020vys}
{Hoy} C.,  {Raymond} V.,  2021, \mn@doi [SoftwareX]
  {10.1016/j.softx.2021.100765}, \href
  {https://ui.adsabs.harvard.edu/abs/2021SoftX..1500765H} {15, 100765}

\bibitem[\protect\citeauthoryear{Hunter}{Hunter}{2007}]{Hunter:2007}
Hunter J.~D.,  2007, \mn@doi [Computing in Science \& Engineering]
  {10.1109/MCSE.2007.55}, 9, 90

\bibitem[\protect\citeauthoryear{{Isaacson}}{{Isaacson}}{1968}]{Isaacson1968}
{Isaacson} R.~A.,  1968, \mn@doi [Physical Review] {10.1103/PhysRev.166.1263},
  \href {https://ui.adsabs.harvard.edu/abs/1968PhRv..166.1263I} {166, 1263}

\bibitem[\protect\citeauthoryear{{KAGRA Collaboration}, {LIGO Scientific
  Collaboration}  \& {Virgo Collaboration}}{{KAGRA Collaboration}
  et~al.}{2019}]{observer_summary}
{KAGRA Collaboration} {LIGO Scientific Collaboration}  {Virgo Collaboration}
  2019, Advanced LIGO, Advanced Virgo and KAGRA observing run plans, \url
  {https://dcc.ligo.org/public/0161/P1900218/002/SummaryForObservers.pdf}

\bibitem[\protect\citeauthoryear{{Kashyap}, {Raman}  \& {Ajith}}{{Kashyap}
  et~al.}{2019}]{Kashyap2019}
{Kashyap} R.,  {Raman} G.,   {Ajith} P.,  2019, \mn@doi [\apjl]
  {10.3847/2041-8213/ab543f}, \href
  {https://ui.adsabs.harvard.edu/abs/2019ApJ...886L..19K} {886, L19}

\bibitem[\protect\citeauthoryear{Kerr}{Kerr}{1963}]{Kerr1963}
Kerr R.~P.,  1963, \mn@doi [Phys. Rev. Lett.] {10.1103/PhysRevLett.11.237}, 11,
  237

\bibitem[\protect\citeauthoryear{Kim, Goobar  \& Perlmutter}{Kim
  et~al.}{1996}]{Kim:1995qj}
Kim A.,  Goobar A.,   Perlmutter S.,  1996, \mn@doi [Publ. Astron. Soc. Pac.]
  {10.1086/133709}, 108, 190

\bibitem[\protect\citeauthoryear{Kluyver et~al.,}{Kluyver
  et~al.}{2016}]{jupyter}
Kluyver T.,  et~al., 2016, in Loizides F.,  Scmidt B.,  eds, Positioning and
  Power in Academic Publishing: Players, Agents and Agendas. IOS Press,
  Netherlands, pp 87--90, \url {https://eprints.soton.ac.uk/403913/}

\bibitem[\protect\citeauthoryear{Kyutoku, Shibata  \& Taniguchi}{Kyutoku
  et~al.}{2010}]{2H}
Kyutoku K.,  Shibata M.,   Taniguchi K.,  2010, \mn@doi [Phys. Rev. D]
  {10.1103/PhysRevD.82.044049}, 82, 044049

\bibitem[\protect\citeauthoryear{{LIGO Scientific Collaboration}}{{LIGO
  Scientific Collaboration}}{2015}]{Voyager_PSD}
{LIGO Scientific Collaboration} 2015, Instrument Science White Paper, \url
  {https://dcc.ligo.org/public/0120/T1500290/002/T1500290.pdf}

\bibitem[\protect\citeauthoryear{Lattimer}{Lattimer}{2012}]{Lattimer2012}
Lattimer J.~M.,  2012, \mn@doi [Annual Review of Nuclear and Particle Science]
  {10.1146/annurev-nucl-102711-095018}, 62, 485

\bibitem[\protect\citeauthoryear{{Lattimer}, {Prakash}, {Masak}  \&
  {Yahil}}{{Lattimer} et~al.}{1990}]{Lattimer1990}
{Lattimer} J.~M.,  {Prakash} M.,  {Masak} D.,   {Yahil} A.,  1990, \mn@doi
  [\apj] {10.1086/168758}, \href
  {https://ui.adsabs.harvard.edu/abs/1990ApJ...355..241L} {355, 241}

\bibitem[\protect\citeauthoryear{{Levin}}{{Levin}}{2007}]{Levin2007}
{Levin} Y.,  2007, \mn@doi [\mnras] {10.1111/j.1365-2966.2006.11155.x}, \href
  {https://ui.adsabs.harvard.edu/abs/2007MNRAS.374..515L} {374, 515}

\bibitem[\protect\citeauthoryear{{MacLeod} \& {Lin}}{{MacLeod} \&
  {Lin}}{2020}]{MacLeod2020}
{MacLeod} M.,  {Lin} D. N.~C.,  2020, \mn@doi [\apj]
  {10.3847/1538-4357/ab64db}, \href
  {https://ui.adsabs.harvard.edu/abs/2020ApJ...889...94M} {889, 94}

\bibitem[\protect\citeauthoryear{McKernan, Ford  \& O’Shaughnessy}{McKernan
  et~al.}{2020}]{McKernan2020}
McKernan B.,  Ford K.,   O’Shaughnessy R.,  2020, Monthly Notices of the
  Royal Astronomical Society, 498, 4088

\bibitem[\protect\citeauthoryear{Messenger, Takami, Gossan, Rezzolla  \&
  Sathyaprakash}{Messenger et~al.}{2014}]{Messenger2014}
Messenger C.,  Takami K.,  Gossan S.,  Rezzolla L.,   Sathyaprakash B.~S.,
  2014, \mn@doi [Phys. Rev. X] {10.1103/PhysRevX.4.041004}, 4, 041004

\bibitem[\protect\citeauthoryear{{Miralda-Escud{\'e}} \&
  {Gould}}{{Miralda-Escud{\'e}} \& {Gould}}{2000}]{Miralda2000}
{Miralda-Escud{\'e}} J.,  {Gould} A.,  2000, \mn@doi [\apj] {10.1086/317837},
  \href {https://ui.adsabs.harvard.edu/abs/2000ApJ...545..847M} {545, 847}

\bibitem[\protect\citeauthoryear{Nitz, Kumar, Wang, Kastha, Wu, Sch\"afer,
  Dhurkunde  \& Capano}{Nitz et~al.}{2021}]{Nitz:2021zwj}
Nitz A.~H.,  Kumar S.,  Wang Y.-F.,  Kastha S.,  Wu S.,  Sch\"afer M.,
  Dhurkunde R.,   Capano C.~D.,  2021, arXiv preprint arXiv:2112.06878

\bibitem[\protect\citeauthoryear{{Norman} \& {Silk}}{{Norman} \&
  {Silk}}{1983}]{Norman1983}
{Norman} C.,  {Silk} J.,  1983, \mn@doi [\apj] {10.1086/160798}, \href
  {https://ui.adsabs.harvard.edu/abs/1983ApJ...266..502N} {266, 502}

\bibitem[\protect\citeauthoryear{Olsen, Venumadhav, Mushkin, Roulet, Zackay  \&
  Zaldarriaga}{Olsen et~al.}{2022}]{Olsen:2022pin}
Olsen S.,  Venumadhav T.,  Mushkin J.,  Roulet J.,  Zackay B.,   Zaldarriaga
  M.,  2022, arXiv preprint arXiv:2201.02252

\bibitem[\protect\citeauthoryear{{Peng} \& {Chen}}{{Peng} \&
  {Chen}}{2021}]{Peng2021}
{Peng} P.,  {Chen} X.,  2021, \mn@doi [\mnras] {10.1093/mnras/stab1419}, \href
  {https://ui.adsabs.harvard.edu/abs/2021MNRAS.505.1324P} {505, 1324}

\bibitem[\protect\citeauthoryear{{Prigozhin} et~al.,}{{Prigozhin}
  et~al.}{2016}]{NICERb}
{Prigozhin} G.,  et~al., 2016, in {den Herder} J.-W.~A.,  {Takahashi} T.,
  {Bautz} M.,  eds,  Society of Photo-Optical Instrumentation Engineers (SPIE)
  Conference Series Vol. 9905, Space Telescopes and Instrumentation 2016:
  Ultraviolet to Gamma Ray. p. 99051I, \mn@doi{10.1117/12.2231718}

\bibitem[\protect\citeauthoryear{Punturo et~al.,}{Punturo et~al.}{2010}]{ET}
Punturo M.,  et~al., 2010, \mn@doi [Classical and Quantum Gravity]
  {10.1088/0264-9381/27/19/194002}, 27, 194002

\bibitem[\protect\citeauthoryear{{Reitze} et~al.,}{{Reitze} et~al.}{2019}]{CE}
{Reitze} D.,  et~al., 2019, in \baas. p.~35 (\mn@eprint {arXiv} {1907.04833})

\bibitem[\protect\citeauthoryear{{Reynolds}}{{Reynolds}}{2014}]{Reynolds2014}
{Reynolds} C.~S.,  2014, \mn@doi [\ssr] {10.1007/s11214-013-0006-6}, \href
  {https://ui.adsabs.harvard.edu/abs/2014SSRv..183..277R} {183, 277}

\bibitem[\protect\citeauthoryear{Saleem et~al.}{Saleem
  et~al.}{2022}]{Saleem:2021iwi}
Saleem M.,  et~al., 2022, \mn@doi [Class. Quant. Grav.]
  {10.1088/1361-6382/ac3b99}, 39, 025004

\bibitem[\protect\citeauthoryear{{Sarin} \& {Lasky}}{{Sarin} \&
  {Lasky}}{2021}]{sarin2021}
{Sarin} N.,  {Lasky} P.~D.,  2021, \mn@doi [General Relativity and Gravitation]
  {10.1007/s10714-021-02831-1}, \href
  {https://ui.adsabs.harvard.edu/abs/2021GReGr..53...59S} {53, 59}

\bibitem[\protect\citeauthoryear{{Schutz}}{{Schutz}}{1986}]{Schutz1986}
{Schutz} B.~F.,  1986, \mn@doi [\nat] {10.1038/323310a0}, \href
  {https://ui.adsabs.harvard.edu/abs/1986Natur.323..310S} {323, 310}

\bibitem[\protect\citeauthoryear{Schutz \& Tinto}{Schutz \&
  Tinto}{1987}]{schutz1987antenna}
Schutz B.~F.,  Tinto M.,  1987, Monthly Notices of the Royal Astronomical
  Society, 224, 131

\bibitem[\protect\citeauthoryear{Soultanis, Bauswein  \& Stergioulas}{Soultanis
  et~al.}{2021}]{Soultanis:2021oia}
Soultanis T.,  Bauswein A.,   Stergioulas N.,  2021, arXiv preprint
  arXiv:2111.08353

\bibitem[\protect\citeauthoryear{Speagle}{Speagle}{2020}]{Speagle:2019ivv}
Speagle J.~S.,  2020, \mn@doi [Mon. Not. Roy. Astron. Soc.]
  {10.1093/mnras/staa278}, 493, 3132

\bibitem[\protect\citeauthoryear{{Syer}, {Clarke}  \& {Rees}}{{Syer}
  et~al.}{1991}]{Syer1991}
{Syer} D.,  {Clarke} C.~J.,   {Rees} M.~J.,  1991, \mn@doi [\mnras]
  {10.1093/mnras/250.3.505}, \href
  {https://ui.adsabs.harvard.edu/abs/1991MNRAS.250..505S} {250, 505}

\bibitem[\protect\citeauthoryear{Tinto}{Tinto}{1987}]{tinto1987antenna}
Tinto M.,  1987, Monthly Notices of the Royal Astronomical Society, 226, 829

\bibitem[\protect\citeauthoryear{{Tsang}, {Dietrich}  \& {Van Den
  Broeck}}{{Tsang} et~al.}{2019}]{Tsang2019}
{Tsang} K.~W.,  {Dietrich} T.,   {Van Den Broeck} C.,  2019, \mn@doi [\prd]
  {10.1103/PhysRevD.100.044047}, \href
  {https://ui.adsabs.harvard.edu/abs/2019PhRvD.100d4047T} {100, 044047}

\bibitem[\protect\citeauthoryear{Unnikrishnan}{Unnikrishnan}{2013}]{LIGO-INDIA}
Unnikrishnan C.~S.,  2013, \mn@doi [International Journal of Modern Physics D]
  {10.1142/s0218271813410101}, 22, 1341010

\bibitem[\protect\citeauthoryear{{Vallisneri}}{{Vallisneri}}{2012}]{Vallisneri2012}
{Vallisneri} M.,  2012, \mn@doi [\prd] {10.1103/PhysRevD.86.082001}, \href
  {https://ui.adsabs.harvard.edu/abs/2012PhRvD..86h2001V} {86, 082001}

\bibitem[\protect\citeauthoryear{Virtanen et~al.}{Virtanen
  et~al.}{2020}]{Virtanen:2019joe}
Virtanen P.,  et~al., 2020, \mn@doi [Nature Meth.] {10.1038/s41592-019-0686-2}

\bibitem[\protect\citeauthoryear{{Wainstein} \& {Zubakov}}{{Wainstein} \&
  {Zubakov}}{1970}]{Wainstein1970}
{Wainstein} L.~A.,  {Zubakov} V.~D.,  1970, {Extraction of Signals from Noise}

\bibitem[\protect\citeauthoryear{{Wang}, {Yan}, {Gao}, {Hu}, {Li}  \&
  {Zhang}}{{Wang} et~al.}{2010}]{Wang2010}
{Wang} J.-M.,  {Yan} C.-S.,  {Gao} H.-Q.,  {Hu} C.,  {Li} Y.-R.,   {Zhang} S.,
  2010, \mn@doi [\apjl] {10.1088/2041-8205/719/2/L148}, \href
  {https://ui.adsabs.harvard.edu/abs/2010ApJ...719L.148W} {719, L148}

\bibitem[\protect\citeauthoryear{{W}es {M}c{K}inney}{{W}es
  {M}c{K}inney}{2010}]{mckinney-proc-scipy-2010}
{W}es {M}c{K}inney 2010, in {S}t\'efan van~der {W}alt {J}arrod {M}illman eds,
  {P}roceedings of the 9th {P}ython in {S}cience {C}onference. pp 56 -- 61,
  \mn@doi{10.25080/Majora-92bf1922-00a}

\bibitem[\protect\citeauthoryear{Zhu, Zhang, Yu  \& Gao}{Zhu
  et~al.}{2021}]{Zhu:2020wrd}
Zhu J.-P.,  Zhang B.,  Yu Y.-W.,   Gao H.,  2021, \mn@doi [Astrophys. J. Lett.]
  {10.3847/2041-8213/abd412}, 906, L11

\bibitem[\protect\citeauthoryear{van~der Walt, Colbert  \& Varoquaux}{van~der
  Walt et~al.}{2011}]{vanderWalt:2011bqk}
van~der Walt S.,  Colbert S.~C.,   Varoquaux G.,  2011, \mn@doi [Comput. Sci.
  Eng.] {10.1109/MCSE.2011.37}, 13, 22

\bibitem[\protect\citeauthoryear{Özel \& Freire}{Özel \&
  Freire}{2016}]{Ozel2016}
Özel F.,  Freire P.,  2016, \mn@doi [Annual Review of Astronomy and
  Astrophysics] {10.1146/annurev-astro-081915-023322}, 54, 401

\makeatother
\end{thebibliography}

\end{document}